\documentclass[longauth]{aa} % for the long lists of affiliations 
\usepackage{graphicx}
\usepackage{txfonts}
\usepackage{natbib}
\bibpunct{(}{)}{;}{a}{}{,}

\usepackage{txfonts}
\begin{document}

   \title{The Gaia-ESO Survey: Empirical determination of the 
   precision of stellar radial velocities and projected rotation velocities
\thanks{Based on observations collected with the FLAMES spectrograph at VLT/UT2 
telescope (Paranal Observatory, ESO, Chile), for the Gaia- ESO Large Public Survey (188.B-3002).}}
%
%   \subtitle{}

   \author{R.~J.~Jackson
          \inst{1}
          \and
          R.~D.~Jeffries \inst{1}
          \and
          J.~Lewis \inst{2}
          \and
          S.~E. Koposov \inst{2,3}
\and
G.~G.~Sacco \inst{4}
\and
S. Randich \inst{4}
\and
G.~Gilmore \inst{2}
\and
M.~Asplund \inst{5}
\and
J.~Binney \inst{6}
\and
P.~Bonifacio \inst{7}
\and
J.~E.~Drew \inst{8}
\and
S.~Feltzing \inst{9}
\and
A.~M.~N.~Ferguson \inst{10}
\and
G.~Micela \inst{11}
\and
I.~Neguerela \inst{12}
\and
T.~Prusti \inst{13}
\and
H-W.~Rix \inst{14}
\and
A.~Vallenari \inst{15}
\and
E.~J.~Alfaro \inst{16}
\and
C.~Allende~Prieto \inst{17,18}
\and
C.~Babusiaux \inst{7}
\and
T.~Bensby \inst{9}
\and
R.~Blomme \inst{19} 
\and
A.~Bragaglia \inst{20}
\and
E.~Flaccomio \inst{11}
\and
P.~Francois \inst{7}
\and
N.~Hambly \inst{10}
\and
M.~Irwin \inst{2}
\and
A.~J.~Korn \inst{21}
\and
A.~C.~Lanzafame \inst{22}
\and
E.~Pancino \inst{20,23}
\and
A.~Recio-Blanco \inst{24}
\and
R. Smiljanic \inst{25}
\and
S.~Van~Eck \inst{26}
\and
N.~Walton \inst{2}
\and
A.~Bayo \inst{27}
\and
M.~Bergemann \inst{2}
\and
G.~Carraro \inst{28}
\and
M.~T.~Costado \inst{16}
\and
F.~Damiani \inst{11}
\and
B.~Edvardsson \inst{21}
\and
E.~Franciosini \inst{4}
\and
A.~Frasca \inst{29}
\and
U.~Heiter \inst{21}
\and
V.~Hill \inst{24}
\and
A.~Hourihane \inst{2}
\and
P. Jofr\'e \inst{2}
\and
C.~Lardo \inst{30}
\and
P.~de~Laverny \inst{24}
\and
K.~Lind \inst{21}
\and
L.~Magrini \inst{4}
\and
G.~Marconi \inst{28}
\and
C.~Martayan \inst{28}
\and
T.~Masseron \inst{2}
\and
L.~Monaco \inst{31}
\and
L.~Morbidelli \inst{4}
\and
L.~Prisinzano \inst{11}
\and
L.~Sbordone \inst{32,33}
\and
S.~G.~Sousa \inst{34}
\and
C.~C.~Worley \inst{2}
\and
S.~Zaggia \inst{15}
         }

   \institute{Astrophysics Group, Keele University, Keele, 
      Staffordshire ST5 5BG, UK\\
              \email{r.j.jackson@keele.ac.uk}
         \and
Institute of Astronomy, University of Cambridge, Madingley Road,
Cambridge CB3 0HA, United Kingdom
\and 
Moscow MV Lomonosov State University, Sternberg Astronomical Institute,
Moscow 119992, Russia
\and
INAF - Osservatorio Astrofisico di Arcetri, Largo E. Fermi 5, 50125,
Florence, Italy
\and
Research School of Astronomy \& Astrophysics, Australian National
University, Cotter Road, Weston Creek, ACT 2611, Australia
\and
Rudolf Peierls Centre for Theoretical Physics, Keble Road, Oxford, OX1
3NP, United Kingdom
\and
GEPI, Observatoire de Paris, CNRS, Universit\'e Paris Diderot, 5 Place
Jules Janssen, 92190 Meudon, France
\and
Centre for Astrophysics Research, STRI, University of Hertfordshire,
College Lane Campus, Hatfield AL10 9AB, United Kingdom
\and
Lund Observatory, Department of Astronomy and Theoretical Physics, Box
43, SE-221 00 Lund, Sweden
\and
Institute of Astronomy, University of Edinburgh, Blackford Hill,
Edinburgh EH9 3HJ, United Kingdom
\and
INAF - Osservatorio Astronomico di Palermo, Piazza del Parlamento 1,
90134, Palermo, Italy
\and
Departamento de F\'{i}sica, Ingenier\'{i}a de Sistemas y Teor\'{i}a de
la Se$\tilde{\rm n}$al, Universidad de Alicante, Apdo. 99, 03080,
Alicante, Spain
\and
ESA, ESTEC, Keplerlaan 1, Po Box 299 2200 AG Noordwijk, The Netherlands
\and
Max-Planck Institut f\"{u}r Astronomie, K\"{o}nigstuhl 17, 69117
Heidelberg, Germany
\and
INAF - Padova Observatory, Vicolo dell'Osservatorio 5, 35122 Padova,
Italy
\and
Instituto de Astrof\'{i}sica de Andaluc\'{i}a-CSIC, Apdo. 3004, 18080,
Granada, Spain
\and
Instituto de Astrof\'{\i}sica de Canarias, E-38205 La Laguna, Tenerife,
Spain
\and
Universidad de La Laguna, Dept. Astrof\'{\i}sica, E-38206 La Laguna,
Tenerife, Spain
\and
Royal Observatory of Belgium, Ringlaan 3, 1180, Brussels, Belgium
\and
INAF - Osservatorio Astronomico di Bologna, via Ranzani 1, 40127,
Bologna, Italy
\and
Department of Physics and Astronomy, Uppsala University, Box 516,
SE-751 20 Uppsala, Sweden
\and
Dipartimento di Fisica e Astronomia, Sezione Astrofisica,
Universit\`{a} di Catania, via S. Sofia 78, 95123, Catania, Italy
\and
ASI Science Data Center, Via del Politecnico SNC, 00133 Roma, Italy
\and
Laboratoire Lagrange (UMR7293), Universit\'e de Nice Sophia Antipolis,
CNRS,Observatoire de la C\^ote d'Azur, CS 34229,F-06304 Nice cedex 4,
France
\and
Department for Astrophysics, Nicolaus Copernicus Astronomical Center,
ul. Rabia\'{n}ska 8, 87-100 Toru\'{n}, Poland
\and
Institut d'Astronomie et d'Astrophysique, Universit\'{e} libre de
Brussels, Boulevard du Triomphe, 1050 Brussels, Belgium
\and
Instituto de F\'isica y Astronomi\'ia, Universidad de Valparai\'iso,
Chile
\and
European Southern Observatory, Alonso de Cordova 3107 Vitacura,
Santiago de Chile, Chile
\and
INAF - Osservatorio Astrofisico di Catania, via S. Sofia 78, 95123,
Catania, Italy
\and
Astrophysics Research Institute, Liverpool John Moores University, 146
Brownlow Hill, Liverpool L3 5RF, United Kingdom
\and
Departamento de Ciencias F\'isicas, Universidad Andr\'es Bello,
Rep\'ublica 220, 837-0134 Santiago, Chile
\and
Millennium Institute of Astrophysics, Av. Vicu\~{n}a Mackenna 4860,
782-0436 Macul, Santiago, Chile
\and
Pontificia Universidad Cat\'{o}lica de Chile, Av. Vicu\~{n}a Mackenna
4860, 782-0436 Macul, Santiago, Chile
\and
Instituto de Astrof\'isica e Ci\^encias do Espa\c{c}o, Universidade do Porto, CAUP, Rua das Estrelas, 4150-762 Porto, Portugal
}
%             \email{c.ptolemy@hipparch.uheaven.space}
%             \thanks{The university of heaven temporarily does not
%                     accept e-mails}

   \date{}

% \abstract{}{}{}{}{} 
% 5 {} token are mandatory
 
  \abstract
   {The Gaia-ESO Survey (GES) is a large public spectroscopic survey at the
European Southern Observatory Very Large Telescope. }
   {A key aim is to
provide precise radial velocities ($RV$s) and projected equatorial
velocities ($v \sin i$) for representative samples of Galactic stars,
that will complement information obtained by the Gaia astrometry
satellite.}
   {We present an analysis to empirically quantify the size and
distribution of uncertainties in $RV$ and $v\sin i$ using spectra from
repeated exposures of the same stars.}
   {We show that the uncertainties
vary as simple scaling functions of signal-to-noise ratio ($S/N$) and $v \sin
i$, that the uncertainties become larger with increasing photospheric
temperature, but that the dependence on stellar gravity, metallicity and 
age is weak.  The underlying uncertainty distributions have extended tails
that are better represented by Student's t-distributions than by normal
distributions.}
   {Parametrised results are provided, that enable
estimates of the $RV$ precision for almost all GES measurements, and
estimates of the $v\sin i$
precision for stars in young clusters, as a
function of $S/N$, $v\sin i$ and stellar temperature. The precision of
individual high $S/N$ GES $RV$ measurements is 0.22-0.26 km/s,
dependent on instrumental configuration.}

\titlerunning{Velocity precision in the Gaia-ESO Survey}
\authorrunning{Jackson et al.}

   \maketitle
%
%________________________________________________________________

\section{Introduction}

The Gaia-ESO survey (GES) is a large public survey programme carried out at
the ESO Very Large Telescope (UT-2 Kueyen) with the FLAMES multi-object
instrument (Gilmore et
al. 2012; Randich \& Gilmore 2013). 
The survey will obtain high- and intermediate-resolution
spectroscopy of $\sim 10^{5}$ stars, the majority obtained at resolving
powers of $R \sim 17\,000$ with the GIRAFFE spectrograph (Pasquini et
al. 2002).  The primary
objectives are to cover representative samples of all Galactic stellar
populations, including thin and thick disc, bulge, halo, and stars in
clusters at a range of ages and Galactocentric radii. The spectra
contain both chemical and dynamical information for stars as faint as
$V \sim 19$ and, when combined with complementary information from the
Gaia satellite, will provide full 3-dimensional velocities and
chemistry for a large and representative sample of stars.
The GES began on 31 December 2011 and will continue for approximately 5
years. There are periodic internal and external data releases, and at the time of writing, 
data from the first 18 months of survey operations have been
analysed and released to the survey consortium for scientific
exploitation -- the ``second internal data release'', known as iDR2. Part of
the same data have also been released to ESO through the second Gaia-ESO phase 3 
and will soon be available to the general community.

The GES data products include stellar radial velocities ($RV$) and
projected rotation velocities ($v\sin i$). 
A thorough understanding of the uncertainties in $RV$ and $v \sin i$ is
an essential component of many aspects of the GES programme. For
instance, the GES data are capable of resolving the kinematics of
clusters and star forming regions, but because the $RV$ uncertainties are
not negligible compared with the observed kinematic dispersion, an
accurate deconvolution to establish intrinsic cluster velocity
profiles, mass-dependent kinematic signatures, net rotation etc.
relies on a detailed knowledge of the $RV$ uncertainties (e.g. Cottaar,
Meyer \& Parker 2012; Jeffries
et al. 2014; Lardo et al. 2015; Sacco et al. 2015).  Searching for
binary members of clusters and looking for outliers in $RV$ space also
requires an understanding of the uncertainty distribution in order to
optimise search criteria and minimise false-positives. Similarly,
inverting the projected rotation velocity distribution to a true
rotation velocity distribution (e.g. Chandrasekhar \& M\"unch 1950;
Dufton et al. 2006) or comparison of 
the rotation velocity distributions of different 
samples requires
an understanding of how uncertainties in $v \sin i$ broaden the
observed distribution and impose a lower limit to the rotation that can
be resolved (Frasca et al. 2015).

These examples illustrate that not only does one wish to know the level
of uncertainty in $RV$ and $v \sin i$ as a function of stellar spectral
type, the spectrum signal-to-noise ratio ($S/N$), the rotation rate and
possibly other variables, but it is also important to understand
whether the uncertainties are normally distributed or perhaps have
extended tails that might be better represented in some other way
(e.g. Cottaar et al. 2014).  The procedures for reducing and analysing
the GES spectra will be fully detailed in forthcoming data release
papers, but ultimately the $RV$s and $v \sin i$ are estimated with a
detailed chi-squared fitting procedure (Koposov et al. in prep and
Sect.~2.3). Fitting uncertainties can of course be computed, but
these are often minor contributors to the overall repeatability of the
measurements and therefore underestimate the total uncertainty.  In
this paper we {\it empirically} determine the uncertainties
and their probability distribution based upon repeated measurements of
the same stars in GES. Our analysis is limited to the $>90$ per cent of
spectra measured with the GIRAFFE spectrograph and deals only with the
{\it precision} of the measurements, rather than their absolute
accuracy. 

In Sect.~2 we describe the GES data and the database of repeat measurements for $RV$
and $v\sin i$ that is available for characterising their
uncertainties. In Sect.~3 we show how the differences in $RV$ and $v \sin i$
measured between repeated observations can be used to determine the
underlying distribution of measurement uncertainty, represented by simple scaling
functions that depend on $S/N$ and $v\sin i$. In Sect.~4 we investigate
how these scaling functions alter with stellar properties. Sect.~5
considers how the measurement uncertainties change for different
observational configurations within GES. In Sect.~6, we conclude and provide
parametric formulae and coefficients that allow an estimation of the $RV$ and $v \sin i$
precision of GES measurements. 

\section{Repeat measurements of radial velocities and projected
  rotation velocities}

\subsection{GES observations}
The GES employs the FLAMES fibre-fed, multi-object instrument, feeding both the
UVES high-resolution ($R \sim 45\,000$) and GIRAFFE intermediate
resolution ($R \sim 17\,000$) spectrographs. More than 90 per
cent of the spectra are obtained with GIRAFFE and we deal only with
these data here. The Medusa fibre system allows the simultaneous
recording of spectra from $\simeq 100$ stars in each pointing. The stars
in a single pointing are usually related by scientific interest (a
cluster or a bulge field etc.) and cover a limited range of brightness
(usually less than a 4 magnitude spread). A further $\simeq 15$ fibres are
normally allocated to patches of blank sky.

The GIRAFFE spectrograph permits the recording of a limited spectral
range and this is selected through the use of order sorting
filters. Eight of these have been used in the GES (HR3, HR5A, HR6, HR10, HR11, HR14A,
HR15N, HR21), each of which records a spectrum over a fixed wavelength
range, although just three filters (HR10, HR15N, HR21) are used for the
large majority of
observations:
\begin{itemize}
	\item Most observations of targets in 
          clusters and star forming regions are made using order
          sorting filter HR15N. The wavelength range of this filter
          (6444-6816\AA) includes both the H$\alpha$ and lithium lines
          and can provide useful information on the effective
          temperature ($T_{\rm eff}$),  gravity ($\log g$), age and
          magnetic activity of the target stars (Lanzafame et al 2015).
	\item Most targets in the halo, bulge and disc
          fields are observed 
         using both filters HR10 and HR21. The main goals here are to
         provide accurate stellar parameters and chemical abundances.
\end{itemize}

GES fields are usually observed in
observation blocks (OBs) comprising two science exposures of equal
duration. In addition, for filters HR10 and HR15N a short "simcal"
exposure is interleaved between the science exposures. The ``simcal''
observation illuminates five dedicated fibres with a Thorium-Argon
(ThAr) lamp,
providing a means of monitoring the wavelength calibration. In the HR21
observations, this role was fulfilled by emission lines in the sky
spectra and no ``simcal'' exposures were performed.

\begin{table}
\caption{Numbers of short and long term repeat GIRAFFE observations of RV and $v\sin i$ used for open clusters with order sorting filter HR15N.}

\begin{tabular}[t]{lcccccc} \hline \hline
Name         & Age    & Ref & \multicolumn{4}{l}{Number of repeat observations}\\
             & (Myr)  &      & \multicolumn{2}{l}{$RV$, $S/N$$>$5}     & \multicolumn{2}{l}{$v\sin i$, $>$5\,km\,s$^{-1}$} \\ 
             &        &         & short & long    &   short & long  \\
             &        &         & term & term    &   term & term  \\\hline                            
Rho Ophiuchi  &    1  & 1 & 222  &    33   & 34   & 2  \\
Chamaeleon I  &    2  & 2 & 617  &    81   & 108  & 22 \\
Gamma Velorum &    6  & 3 &1719  &   523   & 382  & 80 \\
IC4665        &   30  & 4 & 448  &    25   & 43   & 1  \\
NGC2264       &    3  & 5 &2010  &   333   & 717  & 142 \\
NGC2516       &  140  & 6 & 853  &   134   & 266  & 36 \\
NGC2547       &   35  & 7 &1045  &   515   & 321  & 164 \\
NGC6633       &  600  & 8 &1403  &   243   & 103   & 14 \\
Field giants  &  ---   &   & 112  &     9   & 30  & 2 \\\hline

\end{tabular}
\tablefoot{References. (1) Luhman \& Rieke (1999);  (2) Luhman (2007);
  (3) Jeffries et al. (2009); (4) Manzi et al. (2008); (5) Naylor
  (2009); (6) Meynet, Mermilliod \& Maeder (1993); (7) Jeffries \&
  Oliveira (2005); (8) Strobel (1991)}
\label{Clusters}
\end{table}

\subsection{Data reduction}
Full details of the GES GIRAFFE data reduction will be given in a
forthcoming paper (Lewis et al., in prep.). In brief, the raw data
frames are corrected for a bias level using zero exposure bias frames
and the resulting images are divided by normalised daytime tungsten
lamp exposures to remove pixel-to-pixel sensitivity variations. The
multiple spectra in each CCD frame are traced using the tungsten lamp
exposures and then extracted using the optimal algorithm described by
Horne (1986). Given the readout noise and gain of the CCD, 
this algorithm also yields an estimated $S/N$ in the
extracted spectral pixels, 
and it is this estimate that is propagated through subsequent analysis
steps leading to the final reported $S/N$ of the spectra. Extracted
day-time tungsten lamp spectra are used to correct the overall shape of
the spectrum and calibrate the individual transmission efficiencies of
each fibre. The wavelength calibration proceeded in two stages. Deep
exposures of a daytime ThAr lamp are used to define a polynomial
relationship between extracted spectral pixel and wavelength. Then, for
observations using filters HR10 or HR15N the wavelength calibration is
modified by an offset determined from the positions of
prominent arc lines in the night-time ``simcal'' exposures. For
observations using filter HR21 the offset applied to the wavelength
calibration is determined from the position of prominent
emission lines in the sky spectra. Spectra are rebinned into
0.05\,\AA\ pixels using this wavelength solution and sky is subtracted
using a median of the sky spectra corrected for the differing responses
of each fibre.

\subsection{Radial velocity and projected rotation velocity estimates}
The resulting survey spectra are processed and analysed by 
working groups organised in a workflow described by Gilmore et
al. (2012). The $RV$ and $v\sin i$ estimates used in this report are
determined using a pipeline developed by the Cambridge Astronomical
Survey Unit (CASU) which follows the general method described by
Koposov et al. (2011). Details of the pipeline used to analyse the GES
data will be described in a forthcoming paper (Koposov et al. in
preparation). A first pass used a standard cross-correlation method with a
grid of synthetic template spectra at a range of temperatures,
metallicities and gravities (Munari et al. 2005) to give an initial
$RV$ estimate. The second pass used a direct modelling approach that
fits each spectrum with a low-order polynomial multiplied by a template
spectrum, with the $RV$, $v\sin i$, $T_{\rm eff}$, $\log g$, metallicity
and polynomial coefficients as free parameters. The best fit parameter
set is found by chi-squared minimisation with emission lines excluded
from the fitting process. The fitting process is then repeated using a
finer grid to determine optimum values of $RV$ and $v\sin i$ with the
other parameters held constant at their previously determined values.

The chi-squared minimisation yields an estimate of
the uncertainty in the best fit parameters. 
However, in the case of GES data, this under-estimates the 
measurement uncertainty, in
part due to the analysis step where spectra are re-binned but
chiefly due to systematic uncertainties in wavelength calibration
(Jeffries et al. 2014). For this reason an empirical determination of
the measurement precision is preferred; the measurement
uncertainty is estimated by comparing repeated measurements of $RV$ and
$v\sin i$ for the same star.

\begin{table*}
\caption{Log of VLT/Flames observations used in the analysis of $RV$
  and $v\sin i$ measurement precision. The full list is available as Supplementary Material to the on-line version of this paper.}

\begin{tabular}[t]{lllllcccl} \hline \hline
Filter    &       Date     &           UT   &    RA (J2000) &   Dec  (J2000) & Exposure   & Number    & Number&Cluster    \\
          & observation    &                & field centre  & field centre   & time (s)   & exposures & Targets&code     \\ \hline
HR15N    &     2012-02-15  &   03:07:58.00  &   08:10:59.3  &  -47:37:03.5    &     600   &    2   &      111 & gam2vel  \\                             
HR15N    &     2012-02-15  &   03:42:56.00  &   08:09:20.0  &  -47:35:46.3    &     600   &    2   &   112 & gam2vel  \\                       
HR15N    &     2012-02-15  &   04:18:23.00  &   08:07:20.6  &  -47:41:06.0    &     600   &    2   &    81 & gam2vel  \\       
HR15N    &     2012-03-15  &   03:43:29.00  &   11:21:01.7  &  -76:23:40.7    &     600   &    2   &    29 & Cha-I  \\         
HR15N    &     2012-03-16  &   01:39:44.00  &   11:21:01.7  &  -76:23:40.8    &     600   &    2   &    29 & Cha-I  \\\hline

\end{tabular}
\label{log OB S}
\end{table*}

\subsection{Selected data}

To empirically characterise the $RV$ and $v\sin i$ uncertainties and
how they depend on stellar parameters requires a database containing a
large number of repeat observations of the same stars and a broad range
of stellar types and rotational broadening. For these
reasons, and especially to ensure a range of $v \sin i$, we initially
focused on GES data for eight open clusters that were observed using
the HR15N filter.  These clusters have ages in the range 1 to 600\,Myr
(see Table 1), covering both pre-main sequence and main sequence
objects. Only a fraction of the targets in each pointing will be actual
cluster members, but we expect that cluster members will dominate any
subsample of low-mass stars with high $v \sin i$, 
since older field stars are not expected
to rotate quickly.  To provide a sample with older ages
and lower gravities, a field consisting mainly of red giants,
observed on repeated occasions as
part of the GES-CoRoT collaboration, was included.

The data were restricted to observations made with two equal length exposures per
OB. Since this is the usual mode of GIRAFFE observations this leads to
no significant loss of data. Using this standard arrangement simplifies
the analysis and allows two distinct classes of measurement uncertainty
to be identified:

\begin{itemize}
	\item \textbf{Short-term repeats} are where empirical estimates of
          uncertainties are obtained by
          comparing $RV$ and $v\sin i$ values for individual
          targets derived from spectra measured in each of the
          individual exposures {\it within an OB}. The targets are
          observed using the same
          Giraffe fibre in the same configuration and are
          calibrated using the same wavelength solution. In this case the
          uncertainty is expected to be caused primarily by 
          noise in the target spectra and inherent uncertainties in the
          reduction and analysis processes. Any drift
          in wavelength calibration over time, perhaps due to
          temperature or pressure changes, is expected to be small
          since the time delay between exposures is 
          always $<$3000\,s and normally $<$1500\,s; there should also be no
          movement of the fibres and any effects due
          to imperfect scrambling in the fibre or changing hour angle
          (see Sect.~6) should also be small. The assumption is also
          made that any significant velocity shifts due to binary motion on such
          short timescales will be rare enough to be neglected. 

	\item \textbf{Long-term repeats} are where 
          uncertainties are estimated by comparing the mean values of
          $RV$ and $v\sin i$ measured in one OB with those measured for
          the same target {\it in a second OB}, where the fibre allocation and
          configuration on the plate is changed between OBs. In this
          case the empirical uncertainties are due to the combined effects
          of noise in the spectra, the analysis techniques plus any
          external uncertainties
          in the wavelength calibration or possibly differences due to
          the particular fibre used for a target or the hour angle of
          the observation. Binary motion may also contribute to
          any observed velocity shifts. A subset of these long-term
          repeat observations were observations of the same
          star taken on the same night but in a different fibre
          configuration. These are invaluable in assessing the relative
          importance of binaries to the velocity shifts.
 	
The data used in comparing $RV$ measurements were selected to have $S/N>$5 (for the
combined spectra in an OB) and those data used to compare $v\sin i$ have
$S/N>$5 and $v\sin i >5$\,km\,s$^{-1}$. Table 1 shows the number of short
and long term comparisons of $RV$ and $v\sin i$ available for each
cluster.  Table 2 shows the time, date, field centre
co-ordinates, exposure times and numbers of targets for each of the
Giraffe OBs used
in this paper. Values of $RV$, $v\sin i$, $S/N$ and stellar properties
are taken from the iDR2 iteration of analysis of the GES data, first
released by the Cambridge Astronomical Unit to the GES working groups in May 2014 and
subsequently placed in the GES archive at the Wide Field Astronomy Unit at
Edinburgh University\footnote{http//ges/roe.ac.uk/}.

\end{itemize}

\section{Normalised distributions of measurement uncertainty}
Figures~1 and 2 show the general characteristics of the {\it observed} RV
precision, which is defined by the distribution of $E_{RV}= \Delta
RV /\sqrt{2}$, the change in $RV$ between {\it short-term} repeat pairs
of observations for individual targets divided by $\sqrt{2}$. Figure 1 shows
$|E_{RV}|$ for $\sim$8500 short term repeats. There is a strong
dependence on $S/N$ {\it and} $v\sin i$ such that the measurement
precision cannot
be represented by a distribution dependent on just one
of these parameters. Figure~2 compares the distributions of $E_{RV}$ for short-
and long-term repeats. The peak height is reduced and the full width
half maximum (FWHM) is increased for long-term repeats. There is thus
an apparent increase in measurement uncertainty for targets with high
$S/N$ when compared to the precision assessed using short-term repeats of
the same stars.

Our general approach is to divide $E_{RV}$ (and the corresponding
$E_{v\sin i}$) by some
function of the target, signal and spectrograph properties, in order to
identify the underlying normalised distributions of measurement precision. If the
underlying distributions are Gaussian then these normalising functions,
$S_{RV}$ and $S_{v\sin i}$,
would correspond to the standard deviations of $E_{RV}$ and
$E_{v\sin i}$ 
as a function of $S/N$, $v\sin i$ and stellar properties. 
$S_{RV}$ and $S_{v\sin i}$, are used here in a more general
sense in order to normalise the $E_{RV}$ and
$E_{v\sin i}$ distributions to an as yet unknown underlying
distribution which could be non-Gaussian.

Initially, we make the simplifying assumption that the normalising functions
depend only on the $S/N$ and $v\sin i$ of the target star and on the
spectrum resolution and pixel size, which are set by the GIRAFFE order-sorting
filter.

\begin{figure}
	\centering
		\includegraphics[width = 85mm]{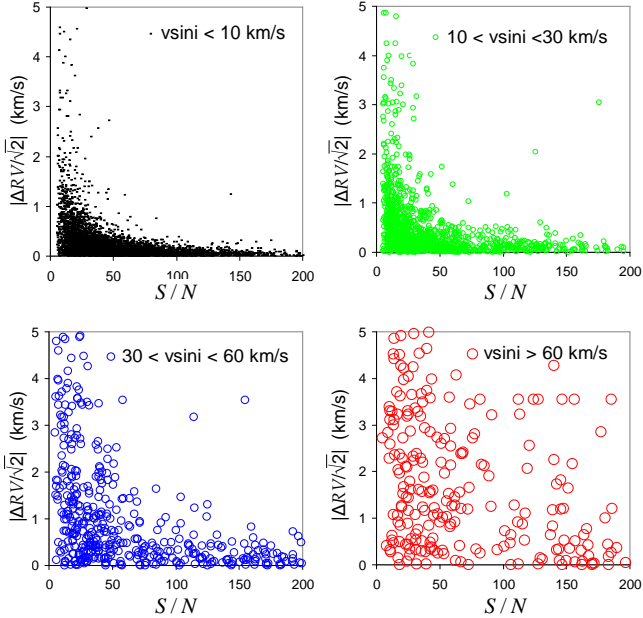}
	\caption{The empirical uncertainty in $RV$ precision
          ($E_{RV}=\Delta RV/\sqrt{2}$) estimated from the change in
          $RV$ between short-term repeat observations of cluster
          targets (see Tables 1 and 2) using order-sorting filter
          HR15N. The size of the symbol indicates the measured value of
          $v\sin i$.}
	\label{fig1}
\end{figure}

\begin{figure}
	\centering
		\includegraphics[width = 80mm]{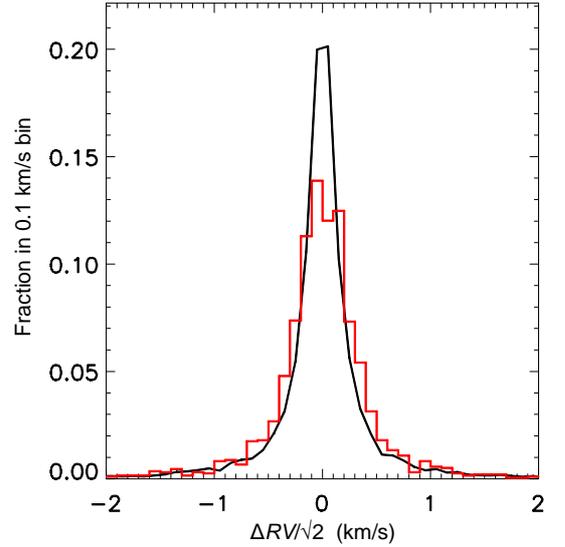}
	\caption{Comparison of the probability density of $E_{RV}$ for
          short- and long-term repeats (see Tables 1 and 2) using
          order-sorting filter HR15N. The black line shows results for
          short-term repeats (i.e. pairs of observations within the 
          same OB). The red histogram shows results for long-term
          repeats (i.e. spectra of the same targets but taken from 
          different OBs where individual targets are allocated to different fibres).}
	\label{fig2}
\end{figure}

\subsection{Normalising functions}
$RV$ and $v\sin i$ are estimated by matching the wavelength offset and
line width of a rotationally broadened template spectrum to the
measured spectrum. To assess the dependency of uncertainty in $RV$ on
$S/N$ and $v\sin i$ it can be shown (see Appendix~A) 
that the distribution of $E_{RV}$
values measured from short term repeats scales approximately according to $W^{3/2}/(S/N)$ where
$W$ is the FWHM of individual lines in a template spectrum, rotationally broadened to match the line width 
 of the measured spectrum.
In this case (also see Appendix A), the $RV$ precision for
short term repeats should scale as 
\begin{equation}
S_{RV,0} = B\frac{(1+([v\sin i]/C)^2)^{3/4}}{S/N}\, ,
\end{equation}
where $C\approx0.895c/R_{\lambda}$, $R_{\lambda}$ is the resolving power of the
spectrograph, $c$ is the speed of light and $B$ is an
empirically determined parameter that will depend on the type of star
being observed. This is consistent with the 
variation of uncertainty in RV with S/N predicted by Butler et al. (1996) for 
photon limited errors.

In the case of long-term
repeats there is an additional contribution to the measurement
uncertainty due to variations in wavelength calibration. This is
independent of $S/N$ and $v\sin i$ and therefore adds a fixed component
$A$ in quadrature to the short term uncertainty such that the
distribution of $E_{RV}$ for long-term repeats scales as; 
\begin{equation}
S_{RV} = \sqrt{A^2+ S_{RV,0}^{2}}\, ,
\end{equation}
where $A$ will be an empirically determined constant and $B$ and $C$
are as defined in Eq.~1.

The \textit{relative} precision of $v\sin i$ used in this paper is
defined as $E_{v\sin i}=\Delta v\sin i/\sqrt{2} \langle v\sin i
\rangle$ (i.e. a {\it fractional} precision), where
$\Delta v\sin i$ is the change between repeat observations and
$\langle v\sin i \rangle$ is their mean value. To find the normalising function
for the $E_{v\sin i}$ distribution we make the assumption that
$W$ increases as a function of $v\sin i$ according to the
rotational broadening function given by Gray (1984) and that the
uncertainty in $W$ varies as $W^{3/2}/(S/N)$. In this case the uncertainty for
short-term repeats (see Appendix A) scales as; 

\begin{equation}
	S_{v\sin i,0}= \beta \frac{(1+([v\sin i]/C)^{2})^{5/4}}{(S/N)\,([v\sin i]/C)^2}\, . 
\end{equation}
Again, a constant term is added in quadrature to account for
additional sources of uncertainty present in the case of long-term
repeats, such that the distribution of $E_{v\sin i}$ scales as;
\begin{equation} 
S_{v\sin i} = \sqrt{\alpha^2+S_{v\sin i,0}^2}\, , 
\end{equation}
where $\alpha$ and $\beta$ will be empirically determined constants and
$C$~is the same function of spectral resolution featured in Eq.~1.

\begin{figure*}
	\centering
		\includegraphics[width = 150mm]{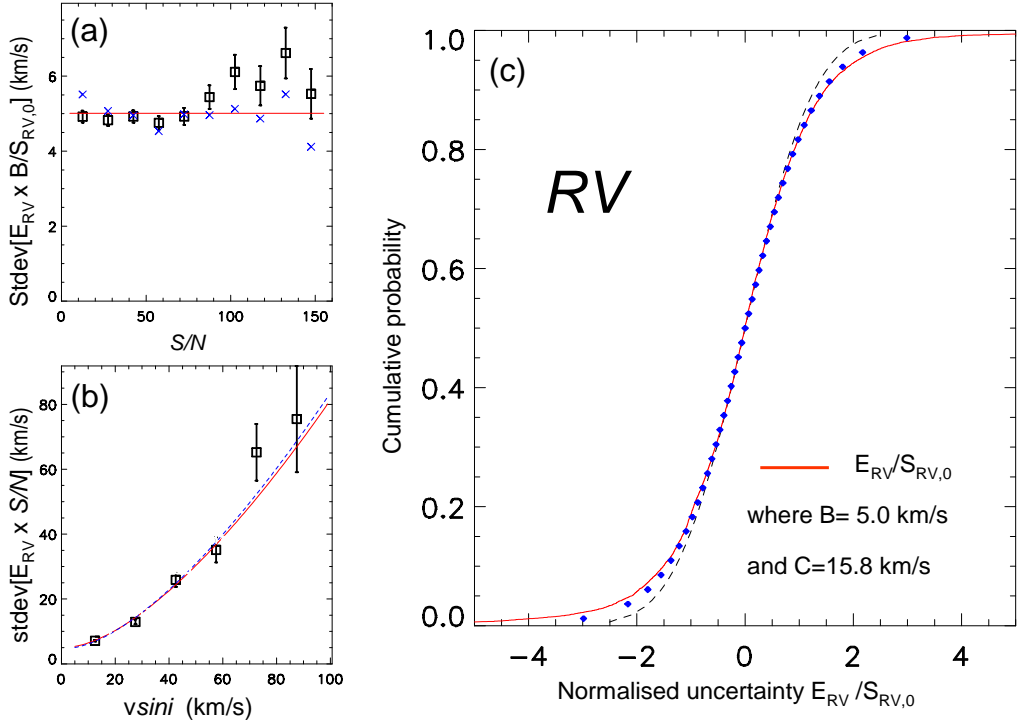} 
	\caption{Analysis of the empirical uncertainty for short-term
          repeat observations of $RV$ using filter HR15N. 
          The solid line in plot (a) shows the variation
          of $E_{RV} \times (S/N)/(1+([v\sin i]/C)^2)^{3/4}$ with $S/N$. The
          horizontal line indicates the value of parameter $B$~ in Eq.~1 fitted to the full
          dataset. Blue crosses show the estimated values $B$ as a function of $S/N$ 
          corrected for the measured variation of $B$ with $T_{\rm eff}$ 
          (see Sect.~4.1 and Table 3). Plot (b) shows the variation of $E_{RV} \times S/N$
          with $v\sin i$. The solid line show the  relationship
          predicted using the theoretical value of 
          $C$~and the value of $B$ from plot (a). The dashed line
          shows a curve of similar functional form using  parameters $B$ and $C$ fitted to 
          the binned data. In plots (a) and (b) the y-axis shows an
          estimate of the standard deviation based on the
          MAD divided by 0.72 (see Sect.~3.2). Plot (c) shows the cumulative probability
          distribution (CDF) of the normalised uncertainty in $RV$ for short-term
          repeats. The red solid line shows results for measured data, the
          dashed line shows the cumulative distribution of a Gaussian
          with unit dispersion,  and the diamond symbols show the cumulative distribution
          function for a Student's t-distribution with $\nu$=6.}
	\label{fig3}
\end{figure*}

\begin{figure*}
	\centering
	\includegraphics[width = 150mm]{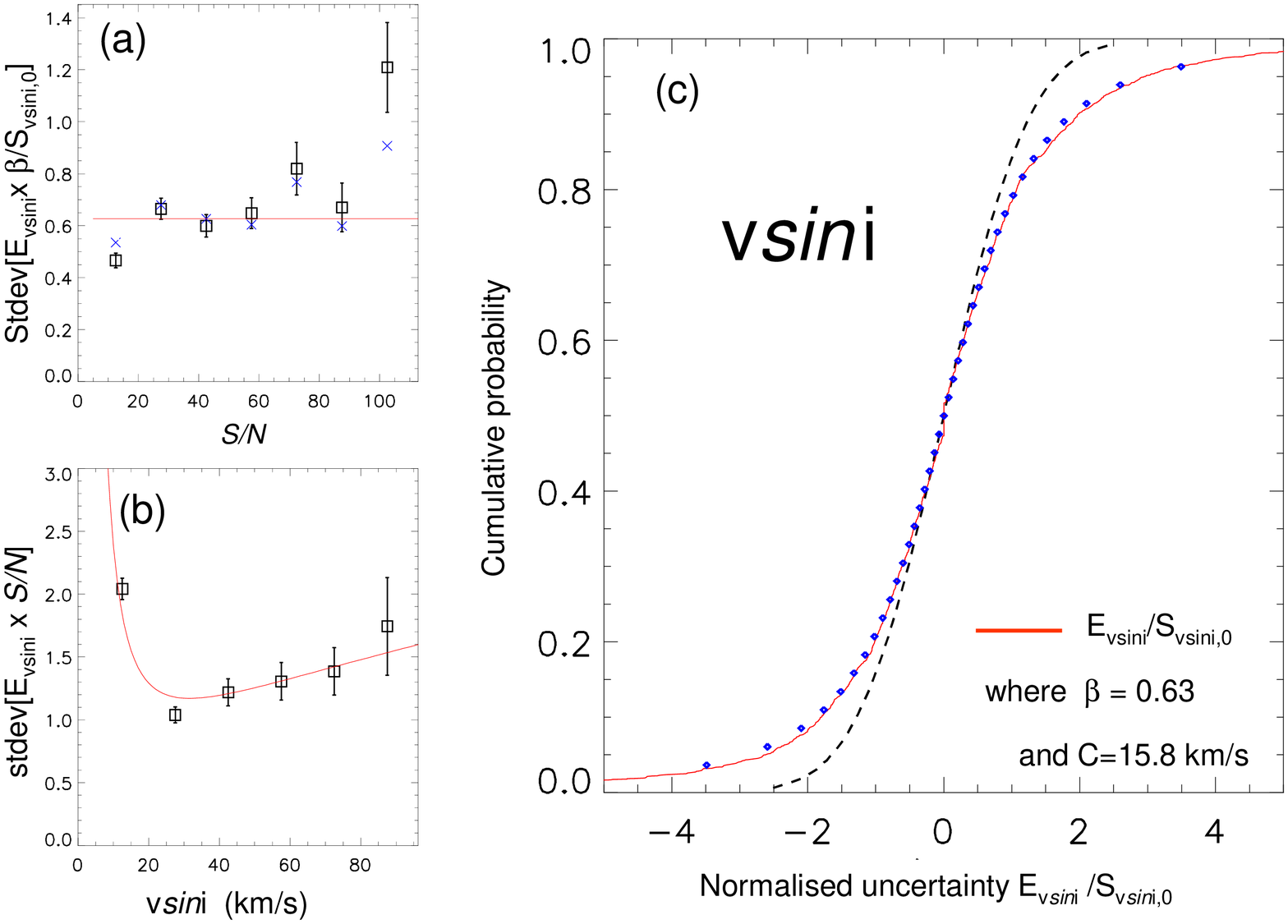}
	\caption{Analysis of the empirical uncertainty for short-term
          repeat observations of $v\sin i$ using order-sorting filter
          HR15N.  Plot (a) shows the variation 
          of $E_{v\sin i} \times (S/N)([v\sin i]/C)^2/(1+([v\sin i]/C)^2)^{5/4}$ with $S/N$ . The
          horizontal line indicates the value of parameter $\beta$~in Eq.~3 fitted to the full
          dataset. Blue crosses show the estimated values $\beta$ as a function of $S/N$ 
          corrected for the measured variation of $\beta$ with $T_{\rm eff}$ (see Sect.~4.4                      
          and Table~3). Plot (b) shows the variation of $E_{v\sin i} \times S/N$
          with $v\sin i$. The solid line show the relationship
          predicted using the theoretical value of 
          $C$~and the value of $\beta$~from plot (a). 
          In plots (a) and (b) the y-axis shows an
          estimate of the standard deviation based on the
          MAD divided by 0.82 (for $\nu$=2, see Sect.~3.3).
          Plot (c) shows the cumulative probability
          distribution (CDF) of the normalised uncertainty in $v\sin i$ for short-term
          repeats. The red solid line shows results for measured data, the
          dashed line shows the cumulative distribution of a Gaussian
          with unit dispersion, 
          and the diamond symbols show the cumulative distribution
          function for a Student's t-distribution with $\nu$=2.}
	\label{fig4}
\end{figure*}

\subsection{Parameters for normalising the $RV$ measurement precision}
Parameters $A$, $B$~and $C$~defining the normalising function $S_{RV}$ are
fitted to match the measured distribution of $E_{RV}$ using a dataset
of 8,429 repeat observations, with $S/N>$5, taken using filter
HR15N. Since we expect (and it turns out) that the distributions of
these quantities are {\it not} Gaussians and have significant
non-Gaussian tails, we choose to use the median absolute deviation
(MAD) to characterise the observed distribution, rather than the square root of
the mean variance which could be heavily biased by outliers. 
An estimate for the standard deviation then follows by noting that the
MAD of a Gaussian distribution is 0.674, such that MAD/0.674 gives an estimate of
the standard deviation. As we shall
see, the distributions more closely follow Student's t-distributions
with $\nu$ degrees of freedom,
for which we determine (by Monte Carlo simulation) the corresponding
corrections of 0.82 for $\nu=2$, 0.77 for $\nu=3$ and 0.72 for $\nu=6$.
Uncertainties in the standard deviations (68 per cent confidence
intervals) as a function of sample size are also estimated using the same Monte Carlo simulations.

Defining $A$, $B$ and $C$ is then done in three steps.
\begin{enumerate}
	\item $B$~is found by finding the MAD of
          $(E_{RV} \times (S/N)/(1+([v\sin i]/C)^2)^{3/4}$, using the
          theoretical value of $C$ determined in Appendix A ($C=15.8$
          km\,s$^{-1}$ for filter HR15N, and see
          step (2) below). Figure~3a shows values of
          $B$~estimated from data in equal bins of $S/N$.  For $S/N<100$
          the average values per bin are close to $B=5.0$\,km\,s$^{-1}$ for
          the full data set. There is more scatter for $S/N>100$ but the
          variation is not excessive considering the larger uncertainties 
          due to the smaller numbers of  data per bin. This indicates that the
          functional form of the normalising function derived in Appendix A 
          is applicable to the GIRAFFE $RV$ data.

	 \item $C$ is then checked by comparing  the curve of 
          $S_{RV,0} \times (S/N)$, calculated using ``empirical''  values of $B$ and $C$
          fitted to the measured values of   $E_{RV} \times (S/N)$
          as a function of $v\sin i$, with the curve predicted using $B$
         and $C$ based on  the  theoretical value of $C$ determined in Appendix A.
         Figure~3b shows that these two curves are very similar for the two methods,
         indicating that the theoretical value of $C$ can be used to predict the 
        scaling of measurement uncertainty in $RV$ with $v\sin i$ . 
         In fact the  uncertainty on the fitted slope is largely due to the 
          relatively small proportion of fast rotating
          stars. For this reason, having confirmed that the data are
          consistent with the theory in Appendix~A, we prefer
          to use the theoretical value of $C$
          rather than an uncertain empirical value.
          The theoretical value for parameter $C$ is a minimum that
          assumes any broadening of the spectral lines beyond the
          spectral resolution is due to rotation. This is reasonable
          for most types of star in the GES, given the modest
          resolution of the GIRAFFE spectra, but if $C$ were underestimated
          then we would over-estimate the increase in measurement
          uncertainty with $v\sin i$ (see Eq. 1).

Figure~3c shows the cumulative distribution function (CDF) of $E_{RV}$ for
short-term repeats normalised with $S_{RV,0}$, together with the CDF of a
unit Gaussian distribution. The distribution of measurement
uncertainties follows the Gaussian distribution over the central region
($-1 \leq E_{RV}/S_{RV,0} \leq 1$), but larger uncertainties are more
frequent than predicted by the Gaussian. The measured distribution of
$E_{RV}/S_{RV,0}$ is better represented by a Student's t-distribution 
with $\nu=6$~degrees of freedom, $\nu$. This value of $\nu$
represents the integer value that provides the best fit to the 
normalised uncertainty of short-term repeats at the 5th and 95th 
percentiles (see Fig.~3). Having determined this, steps (1) and (2) are iterated,
dividing the MADs by the appropriate factor of 0.72 (for a Student's
t-distribution with $\nu=6$)  to estimate a true
standard deviation and produce the final results.

	\item The value of $A$ that is added in quadrature to $S_{RV,0}$ 
	        is set to $A=0.25\pm0.02$\,km\,s$^{-1}$. 
	        This value is chosen so that the normalised
          CDF of observed $E_{RV}$ found from long-term repeats,
          $E_{RV}/S_{RV}$, matches the normalised distribution 
          of uncertainty from short-term
          repeats ($E_{RV}/S_{RV,0}$), but {\it only between the upper and
          lower quartiles}. We choose only to match this range because
          the tails of the distribution are {\it expected} to be
          different owing to the likely presence of binaries. We show
          in Sect.~4.3 that this assumption is justified because the
          distribution of $E_{RV}/S_{RV}$ for those ``long-term'' repeats
          where the repeat observations were taken on the same
          observing night is indistinguishable from that of
          $E_{RV}/S_{RV,0}$ for short-term repeats both in the core and
          the tails of the distribution.
 
          The value of $A$ defines the minimum level
          of uncertainty that can be achieved for GES spectra with high $S/N$.
	
\end{enumerate}

Figure 3a shows an increase in the estimated value of $B$ for $S/N > 100$. 
This does not significantly affect the estimate of parameters $A$, $B$ and $C$ described 
above but does reflect the variation of $B$ with stellar properties. Lower $S/N$ bins contain 
a mix of stars such that variations of $B$ with stellar properties average out. 
However the smaller samples in the high $S/N$ bins contain a higher fraction of stars 
with larger $T_{\rm eff}$ and, as we show in Sect.~4.1, $B$ increases with $T_{\rm eff}$. 
The blue crosses in Fig.~3a show $T_{\rm eff}$-corrected values $B$ as a function of $S/N$ , 
using the values discussed in Sect. 4.1 and reported in Table 3. These show a more uniform variation of $B$ with $S/N$.

\subsection{Parameters for normalising the $v\sin i$ precision}

Constants $\alpha$, $\beta$ and $C$~that define the normalising function
$S_{v\sin i}$ are fitted to match the measured distribution
of $E_{v\sin i}$ for  a subset of the data comprising 2004
observations with $v\sin i>5$\,km\,s$^{-1}$. Again, parameters
are evaluated in three steps with the MAD being  used to estimate 
the true standard deviations and the analysis being iterated once the true
distribution of $E_{v\sin i}/S_{v\sin i}$ is known. 

\begin{enumerate}
	\item   First $\beta$~is found by determining the MAD of
          $E_{v\sin i}(S/N)([v\sin i]/C)^2/(1+([v\sin i]/C)^2)^{5/4}$, using the
           theoretical value of $C$ determined in Appendix A. 
          The variation of the uncertainty with $S/N$ shows some scatter
          (see Fig.4a) and consequently there is an
          $\pm$8 per cent uncertainty in the estimated value of $\beta$.

  \item   $C$ is then checked by comparing  the measured values of $E_{v\sin i} \times (S/N)$~
          as a function of $v\sin i$ with the curve predicted using $\beta$
         and $C$ based on  the  theoretical value of $C$ determined in Appendix A.
         Figure~4b shows reasonable agreement between the semi-empirical curve 
         and the measured data indicating that a scaling function
         of the form $S_{v\sin i} $ using the theoretical value of $C$ can be 
         used to predict the variation of measurement uncertainty
         with $S/N$ and  $v\sin i$.
         
      Figure 4c shows the CDF of $E_{v\sin i}/S_{v\sin i, 0}$ for short-term repeats. This shows a more
      pronounced tail than the normalised distribution of $E_{RV}$ precision
      (Fig.~3c) such that a broader Student's t-distribution with $\nu$=2
     is a better fit to the CDF between the 5th and 95th percentiles.

    \item Finally, the value of $\alpha$ that represents the effect of wavelength 
    uncertainty for long term repeats is found by matching the
    normalised $E_{v\sin i}/S_{v\sin i}$ distribution from 463 long-term
    repeats with the equivalent distribution for the short-term repeats 
    between the upper and lower quartiles, giving  $\alpha =0.047 \pm 0.003$. 
    This corresponds to the minimum {\it fractional} uncertainty in $v\sin i$ 
    that can be obtained from GES spectra with high $S/N$ and large $v\sin i$. 
    This optimum result is most readily achieved in spectra 
    with $v\sin i = 2C$ (i.e. 31\,km\,s$^{-1}$). Figure 4b shows that, for a given (S/N), 
    fractional uncertainties rise at both higher and lower values of $v\sin i$, and rise 
    drastically for $v\sin i < 10$\,km\,s$^{-1}$ due to the limited spectral
    resolution.

\end{enumerate}

Figure~4a shows an increase in the estimated value of $\beta$ for $S/N > 100$ due to the 
higher proportion of hotter stars in this bin. This variation is reduced when the estimated 
value of $\beta$ is corrected for the measured variation of $\beta$ with $T_{\rm eff}$ 
discussed in Sect.~4.4 and reported in Table~3.

\section{The Effect of Stellar Properties}
In Sect.~3 the constants defining the normalising functions $S_{RV}$ and
$S_{v\sin i}$ were estimated by fitting data from an inhomogeneous set
of stars. The
values obtained represent average values. In
this section we determine how these ``constants''
vary with stellar properties, in particular 
$T_{\rm eff}$, gravity, metallicity, and age. We make the simplifying assumption that
uncertainties in $RV$ and $v\sin i$ scale with $S/N$ and $v\sin i$ as
described in the last section and that only the parameters $B$~in Eq.~1
and $\beta$ in Eq.~3 depend on stellar properties. This follows because
parameters $A$ and $\alpha$~represent uncertainties due
to changes in wavelength calibration with time and fibre configuration,
and parameter $C$~should depend only on the spectral resolution (Eq.~A4).

\begin{figure}
	\centering
		\includegraphics[width = 80mm]{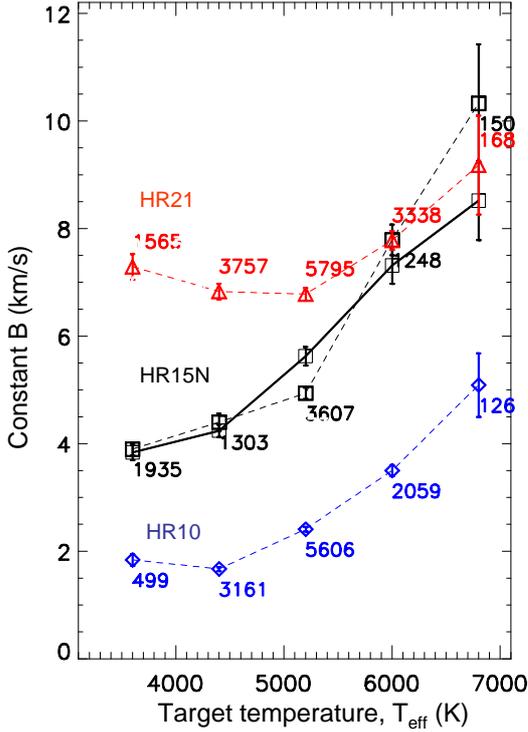}
	\caption{Variation of parameter $B$~of the scaling function for
          uncertainty in $RV$~($S_{RV}$) with effective
          temperature. The solid line shows results for filter HR15N as
          a function of $T_{\rm eff}$ (see section
          4.1). Dashed lines show results for filters HR10, HR15N and
          HR21 as a function of the temperature of the template
          spectrum fitted in the CASU pipeline (see Sect.~2.3). 
          Numbers equal the sample size per bin.}
	\label{fig5}
\end{figure}

\subsection{Variation of $S_{RV}$ with effective temperature}
Values of $T_{\rm eff}$ determined from an analysis of the iDR2
spectra are available in the GES archive for 75 per cent of
HR15N targets considered in this paper. It is labelled {\it Teff} in the
archive. The $E_{RV}$ values
are divided between 5 evenly spaced bins of temperature between 3000\,K and 7000\,K
and analysed as described in section~3.2. The results in Fig.~5
show a slow increase of $B$~with temperature for
$T_{\rm eff}<5200$\,K such that $B$~is within $\pm$10 per cent of the mean
value in Fig.~3b. However, above 5200\,K, $B$~increases rapidly with temperature
to twice its mean value at $T_{\rm eff} \sim  7000$\,K.

The dashed lines in Fig.~5 also show results plotted as a function of
the  "template" temperature (known in the GES archive as
{\it logTeff}). This is the logarithm of the temperature of the best-fit
synthetic spectrum that was used to determine $RV$
and $v\sin i$ in the pipeline. This is likely to be less accurate
than the $T_{\rm eff}$ derived from a full spectral analysis, but 
a key advantage is that it is available
for all iDR2 targets with a $RV$ and $v\sin i$. In fact, the $B$~values
estimated using the "template" temperature have a very similar trend with
$T_{\rm eff}$ and so may be used directly to estimate
temperature-dependent  values of $B$~and $S_{RV}$ where these are required.

\begin{figure}
	\centering
		\includegraphics[width = 85mm]{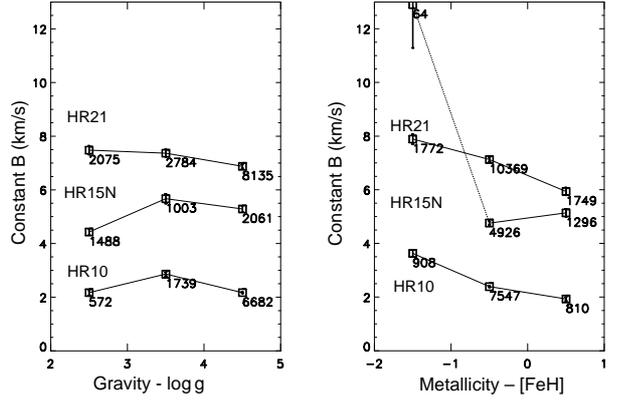}
	\caption{Variation of parameter $B$~of the scaling function for
          uncertainty in $RV$~($S_{RV}$) with $\log g$ and
          metallicity for order sorting filters, HR10, HR15N and HR21. 
          (a) shows the value of $B$~determined for data in
          three equal bins of $\log g$. Numbers indicate the sample size. 
          (b) shows the value of $B$~determined for data in three equal bins
          of [Fe/H]. For filter HR15N only the values shown in the upper two bins are 
          reliable due to the low number of targets with [Fe/H]$<$-1.}
	\label{fig6}
\end{figure}

\begin{figure}
	\centering
		\includegraphics[width = 70mm]{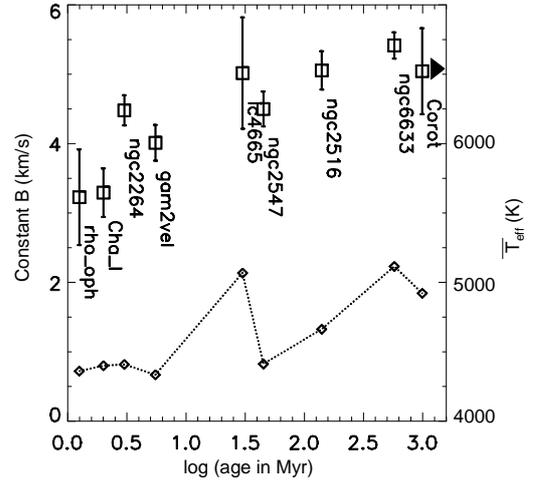}
	\caption{Variation of parameter $B$~of the scaling function for
          uncertainty in $RV$~($S_{RV}$) with target age for observations 
          with order sorting filter HR15N. Square symbols show the  
          value of $B$~for targets identified as possible cluster
          members from their $RV$ versus the nominal age of the cluster
          (see Table 1). No selection by $RV$ is made for NGC6633 or
          COROT and we assume the stars have mean ages $>1$~Gyr. The dotted line 
          indicates the median value of $T_{\rm eff}$ (right hand axis
          values) for members identified in each cluster.}
	\label{fig7}
\end{figure}

\begin{figure}
	\centering
		\includegraphics[width = 70mm]{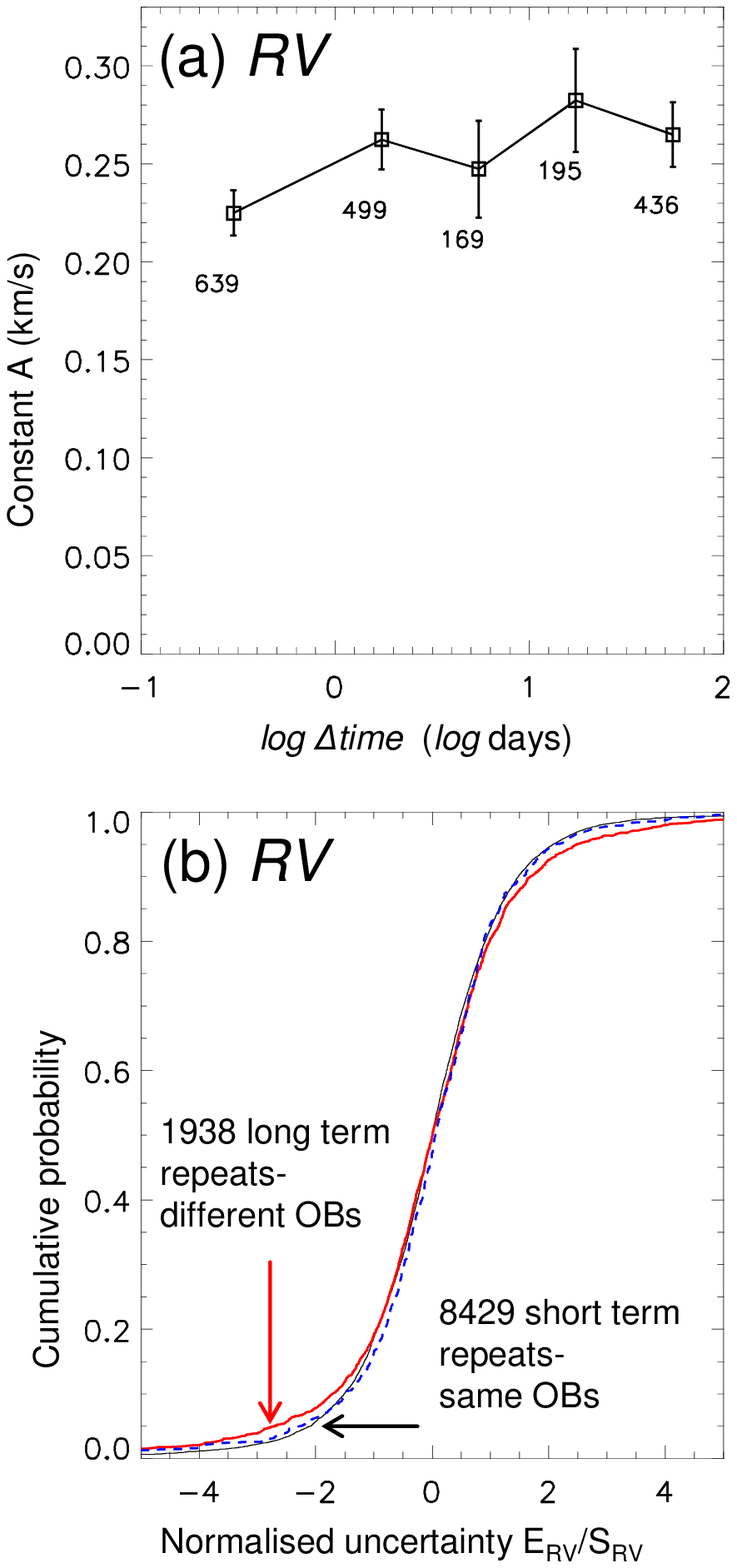}
	\caption{The dependence of the scaling function for uncertainty
          in $RV$~($S_{RV}$) as a function of time between
          observations. Plot (a) shows how scaling parameter
          $A$ varies with time between repeat observations. Numbers indicate
          size of the sample used to determine A. Plot (b) shows the CDFs
          of the normalised distribution of $RV$ precision for
          short-term repeats (black line), long-term repeats (red line)
          and long term repeats where the observations were taken on the
          same observing night (blue dashed line). The CDFs for the
          short-term repeats and the long-term repeats within a night
          are indistinguishable using a two-tailed Kolmogorov-Smirnov test.}
	\label{fig8}
\end{figure}

\subsection{Variation of $S_{RV}$ with gravity, metallicity and age}
Values of $\log g$  and [Fe/H] (labelled as {\it logg} and {\it FeH} in the GES archive) 
obtained from a detailed spectral analysis by the GES working groups
are presently available for about 75 percent of the targets observed with order sorting filter HR15N. Analysing these data in bins of $\log g$ (see Fig.~6a) shows
only a $\sim$25 percent change in the estimated value of parameter
$B$~over a 2 dex range in $\log g$. Analysis in bins of [Fe/H] (see Fig.~6b) shows
a similarly small change in $B$ with metallicity over the range of
metallicities -1$<$[Fe/H]$<$1. Below this, the estimated value of
parameter $B$ appears to increase sharply with decreasing [Fe/H] but in
truth there are too few data points for filter HR15N with [Fe/H]$<$-1
to estimate parameter $B$ with any degree of accuracy. We confirmed
that any variation seen in Fig.~6 is {\it not} due to differences in
temperature  -- the median values of {\it
  logTeff} are very similar in all binned subsamples.

Although the fundamental
cause of any variation of $RV$ precision with age 
would likely be due to the evolution of $\log g$ in
pre-main sequence stars, it is nevertheless important to confirm that
the prescription for calculating $RV$ precision is valid at all ages,
since studying the dynamics of young clusters is a key GES objective. 
Figure~7 shows the variation of $B$~with stellar age. The adopted ages
for cluster stars are those given in Table 1. For this plot, we attempted to
separate genuine cluster members from field objects
by selecting according to $RV$.  For most cluster datasets there was
a clear $RV$ peak corresponding
to the cluster, so cluster members were selected from a range
$\pm$5\,km\,s$^{-1}$ either side of this peak with little contamination.
However, no selection by $RV$ was made for the COROT sample or
for the cluster NGC6633, since neither showed a clear peak in their $RV$
distributions. We assume these datasets contain mostly older ($>1$~Gyr) field
stars. Figure~7 shows in any case that there {\it is} a weak dependence
of $B$ on age. However, it can be seen that this small variation is directly
linked to the decreasing median temperatures of the cluster
samples at younger ages.

\subsection{Variation of $S_{RV}$ with time between observations}
In our model of $RV$ uncertainty we assume that 
$A$~represents some additional uncertainty arising from random changes
in wavelength calibration with time and the effects of changes in fibre
allocation. We fitted $A$, using the interquartile range of the uncertainty
distribution in long-term repeats, in an effort to avoid modelling tails that might be
due to binary motion. This simplifying assumption can be tested by plotting
values of $A$~determined for samples with increasing time differences between
observations. Figure~8a shows that the value of $A$~depends
only weakly on the time between observations, increasing from $0.23\pm 0.02$\,km\,s$^{-1}$
for measurements made in different configurations on the same night to
$0.26\pm 0.02$\,km\,s$^{-1}$ for intervals of up to 100 days between
observations. This confirms that the $A$ value is not unduly influenced
by any binaries in the sample.

The effect of binaries is far more apparent in the tails of the distributions.
Figure~8b compares CDFs of the normalised distribution
of measurement uncertainty derived from the change in $RV$ between
short-term repeats, normalised with $S_{RV,0}$ (Eq. 1), with (i) all the 
long-term repeats, 
with uncertainties normalised to $S_{RV}$ (Eq. 2); (ii) a
separate distribution of $E_{RV}/S_{RV}$ for just those long-term
repeats where the repeat observations were on the same night
(nullifying the effects of all but the
rarest, short-period binaries).
By design, the three CDFs are very close in the interquartile range;
but while the CDF for long-term repeats has a more
pronounced tail, better described by a Student's t-distribution
with $\nu$=3,
the long-term repeats within a night are indistinguishable (with a
Kolmogorov-Smirnov test) 
from the short-term repeats, following a Student's t-distribution with $\nu=6$.
This is consistent with a fraction of the sample being
binary stars that show genuine $RV$ changes between
observations on timescales longer than a day. It also justifies an
assumption that the true uncertainties in a single RV measurent are
best represented by the $\nu=6$ Student's t-distribution multiplied by
$S_{RV}$ as given by Eq.~2.

\begin{figure}
	\centering
		\includegraphics[width = 70mm]{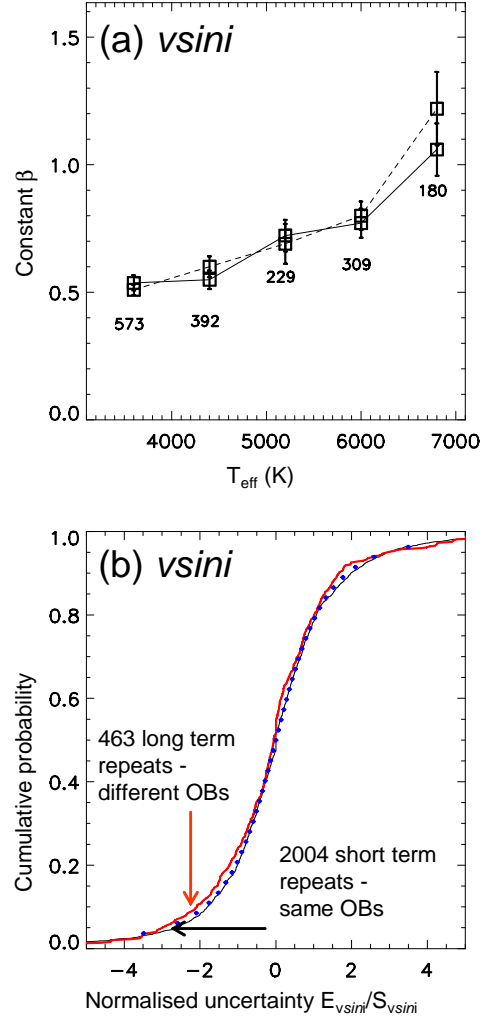}
	\caption{Variation of the scaling function for $v\sin i$ with
          temperature and time between observations. Plot (a) shows how
          the $\beta$ parameter in Eq.~3 varies with
          $T_{\rm eff}$. The solid line shows results using $T_{\rm eff}$ from
          a detailed spectral analysis; the dashed line shows the results
          using the "template" temperature (see Sect.~4.4). Labels indicate
          the sample size per bin. Plot (b) shows the CDF of the
          normalised distributions of $v\sin i$ uncertainty for short and
          long term repeats. Also shown (as small diamonds) is a
          Student's t-distribution
          with $\nu$=2.}
\label{fig9}
\end{figure}

\subsection{Variation of $S_{v\sin i}$ with temperature and time between observations}
In Fig.~9a we show how $\beta$, the parameter in the scaling function governing
$v\sin i$ precision (see Eq.3), depends on stellar
temperature and the time between observations . The data were divided into 5 equal
bins of temperature. Results are shown using the temperature derived from
detailed spectral analysis ({\it Teff}) and the best-fitting "template" temperature
({\it logTeff}). There appears to be little variation with $T_{\rm
  eff}$ below 6000\,K using either temperature estimate, but like
the parameter $B$ governing
$RV$ precision, there is a rapid growth in $\beta$ for hotter stars -- by about a
factor of 2 at $T_{\rm eff} \simeq 7000$\,K.

Figure~9b compares the CDF of the normalised measurement
precision in $v\sin i$ for short- and long-term repeats. There is much
less difference between these CDFs than that found between the short- and
long-term repeat estimates of $RV$ precision. This is not unexpected since measurements of
$v\sin i$ should be much less effected by binarity. A Student's t-distribution with 
$\nu=2$ fits either the short- or long-term repeat CDFs equally well.

There are too few stars in our sample with $v \sin i>5$ km\,s$^{-1}$ 
for a detailed
investigation of how $\beta$ might vary with age, $\log g$ or
metallicity subsets.

\begin{figure*}
	\centering
		\includegraphics[width = 170mm]{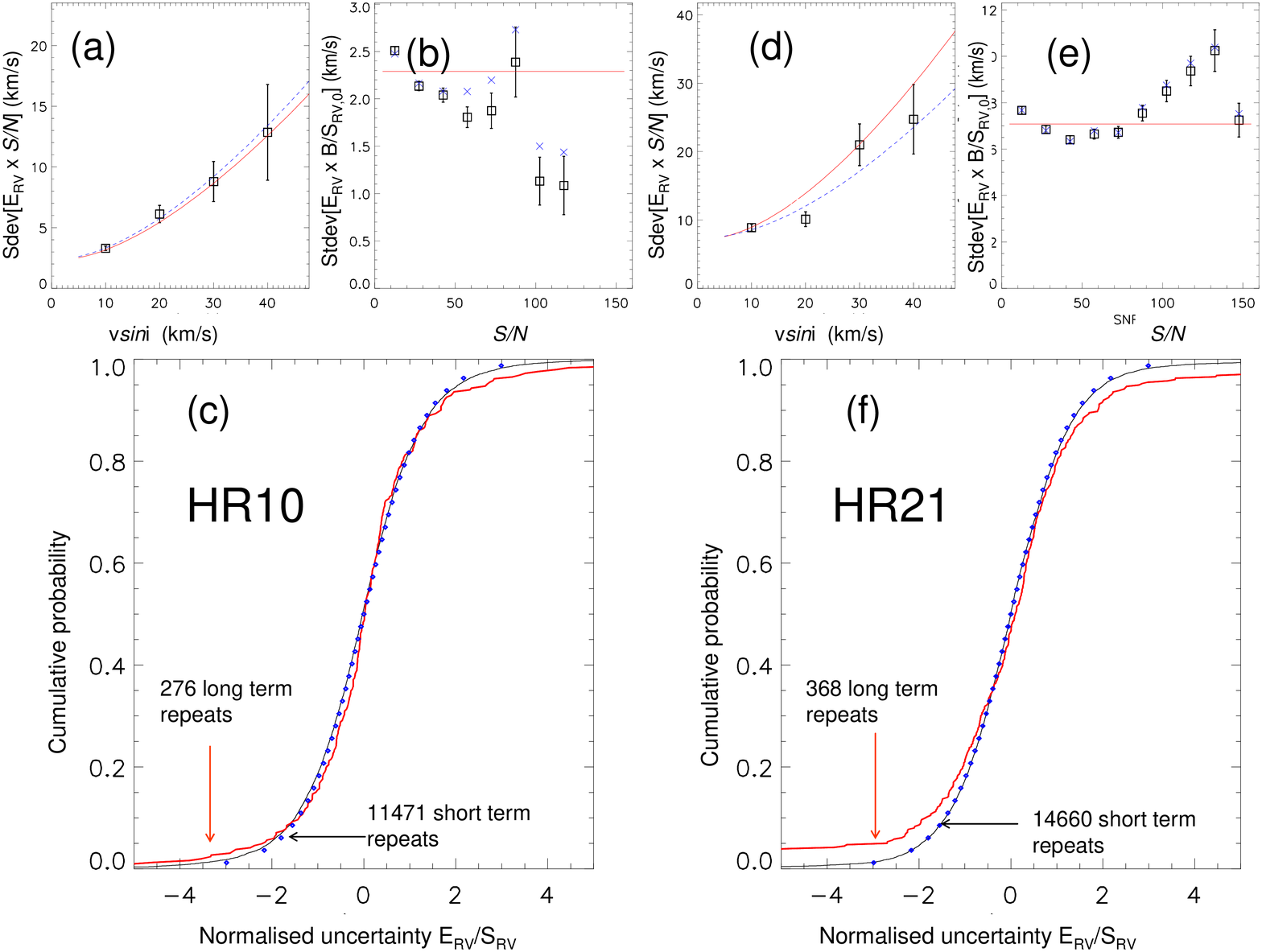}
\caption{Analysis of the empirical uncertainty for short term repeat
  observations of $RV$ using filter HR10 and HR21. Plots a and d show
  the variation of $E_{RV}$ with $v\sin i$. Red lines show the
  curves predicted using the model value of $C$~(see Eq.~A4). The dashed lines
  shows the curves predicted using values of $B$ and $C$ fitted to the binned data.
  Plots b and e show the variation in the estimated value of parameter $B$ in Eq.~1 with    
  $S/N$. Red lines show values of $B$ fitted to the full data set for each filter. 
  Blue crosses show the estimated values $B$ as a function of $S/N$ 
  corrected for the measured variation of $B$ with $T_{\rm eff}$ 
  (see Table 3). Plots c and f shows
  the normalised uncertainty for short term repeats and long term
  repeats. The black curve shows the CDF for short-term repeats. The
  red line shows the CDF for long-term repeats. Blue diamonds show a
  Student's t-distribution with $\nu=6$, which matches the distribution
  for short-term repeats well.}
\label{fig10}
\end{figure*}

\begin{table}
\caption{Constants describing the scaling function of measurement precision in $RV$ and $v\sin i$ as a function of $S/N$ and $v\sin i$ (see Eqns.~2 and 4).}
\begin{tabular}[t]{lccc} \hline \hline
\multicolumn{4}{l}{Characteristics of order sorting filter}\\

Filter                     & HR10  & HR15N         & HR21  \\ 
Mean $\lambda$~(\AA)     & 5470  & 6630          & 8728  \\ 
Resolution                 & 19800 & 17000         & 16200 \\
Range $\lambda$~(\AA)    & 270   & 370           & 504   \\\hline
\multicolumn{4}{l}{Parameters defining the scaling function $S_{RV}$ (Eq.~3)}\\
A\,(km\,s$^{-1}$)               & 0.22$\pm$0.02 & 0.25$\pm$0.02 & 0.26$\pm$0.02\\
C\,(km\,s$^{-1}$)               & 13.6          & 15.8          &  16.6 \\\hline
\multicolumn{4}{l}{Average value of $B$ for the mix of stars analysed in this paper}\\
B\,(km\,s$^{-1}$)               & 2.3           &  5.0          &  7.1 \\
\multicolumn{4}{l}{Variation of $B$ with template temperature (section~4.1) }\\
B (3200-4000\,K)        & 1.8$\pm$0.2   & 3.9$\pm$0.1   &7.3$\pm$0.2\\
B (4000-4800\,K)        & 1.7$\pm$0.1   & 4.4$\pm$0.2   &6.8$\pm$0.2\\
B (4800-5600\,K)        & 2.4$\pm$0.1   & 4.9$\pm$0.1   &6.8$\pm$0.1\\
B (5600-6400\,K)        & 3.5$\pm$0.2   & 7.8$\pm$0.3   &7.8$\pm$0.2\\
B (6400-7200\,K)        & 5.1$\pm$0.6   &10.3$\pm$1.1   &9.2$\pm$1.0\\\hline
\multicolumn{4}{l}{Parameters defining the scaling function $S_{v\sin i}$ (Eq.~4)}\\
$\alpha$        &     ----     & 0.047$\pm$0.005    & ---- \\
C\,(km\,s$^{-1}$)               &     ----     & 15.8     & ---- \\\hline
\multicolumn{4}{l}{Average value of $\beta$ for the mix of stars analysed in this paper}\\
$\beta$           &    & 0.63      & \\
\multicolumn{4}{l}{Variation of $\beta$ with template temperature (section~4.4) }\\
$\beta$ (3200-4000\,K)    & ----  & 0.51$\pm$0.02      & ---- \\
$\beta$ (4000-4800\,K)    & ----   & 0.60$\pm$0.04      & ---- \\
$\beta$ (4800-5600\,K)    & ----   & 0.69$\pm$0.08      & ---- \\
$\beta$ (5600-6400\,K)    & ----   & 0.80$\pm$0.06      & ---- \\
$\beta$ (6400-7200\,K)    & ----   & 1.22$\pm$0.15      & ---- \\\hline
\multicolumn{4}{l} {* Where C is calculated from Eq.~4 assuming a limb darkening }\\
\multicolumn{4}{l} { coefficient $u=0.6$~ (Claret Diaz-Cordoves \& Gimenez 1995).}
\\
\end{tabular}
\label{parameters}
\end{table}

\section{Measurement uncertainties using different instrumental configurations}
So far the analyses have been restricted to observations with the HR15N
order-sorting filter. In this section we consider how these results can
be extended to the other GES observational setups. We used all of the
``GES\_MW'' (GES Milky Way Programme) fields, consisting of more than
20,000 $RV$ measurements from individual 
spectra taken with the HR10 and HR21 filters.
Unfortunately there are too few
measurements through these filters with $v\sin i>5$ km\,s$^{-1}$ to constrain  
the $v \sin i$ precision in the same way that was done for HR15n observations.

These precision of the HR10 and HR21 $RV$ measurements were compared with those
predicted using the
simple model described in Appendix A. For this comparison it is
assumed that:
\begin{itemize}
	\item The uncertainty in $RV$ precision scales as $S_{RV}$ (See Sect.~3.1)
	\item Parameter $C$~characterising the dependence of $RV$ uncertainty on $v\sin i$ 
     depends on the spectral resolution as $0.895 c/R_{\lambda}$ (see Eq.~A4).
	\item Parameter $A$~that determines the difference in $RV$
          precision for short- and long-term repeats corresponds to a
          displacement of the spectrum on the detector in the
          dispersion direction, measured in pixels, rather than a fixed velocity
          difference. The number of physical CCD pixels contributing
          each spectrum along the dispersion direction is 4096, so
          $\delta_{pix} = 4096
          \overline{\lambda} A/c \Delta \lambda$ where $\Delta \lambda$
          is the wavelength range of the filter (see Table 3). In the
          case of filter HR15, $A=0.25$\,km\,s$^{-1}$ corresponds to
          $\delta_{pix}=0.061$ pixels, which we will assume is the same for
          the other filters.

\end{itemize}
Only constant $B$~has to be found, and this can be done using the
distribution of $E_{RV}$ found from short-term
repeats. This allows the measurement precision of $RV$ for a given
filter to be estimated even when there are no long-term repeat
measurements to make an independent
empirical analysis.  In the case of filters HR10 and HR21 it turns out
that there are enough long-term repeat measurements, albeit over a
restricted range of $v\sin i$ values, to test this hypothesis.

Figure~10 shows an analysis
for all field stars that were observed in the GES\_MW
fields. Figures 10a and d show the variation of the standard deviation of  
$E_{RV}\times S/N$ with $v\sin i$. Data with large
$v\sin i$ values are few; therefore the error bars become large with
increasing $v \sin i$. Even so, the curve corresponding to the
value of $B$ evaluated for the full data set using the value of
$C$ predicted from Eq.~A4 is consistent with the empirically 
measured uncertainties.

Figures 10b and e show how the standard deviation of  
$(E_{RV})(S/N)/(1+([v\sin i]/C)^2)^{3/4}$ (estimated using the MAD) varies with $S/N$.
For $S/N<100$ both plots show reasonable agreement (within 10 percent) between the measured data 
and the line showing a single value of $B$ evaluated for the full data set 
using theroretical $C$. Agreement is less good for 
data with $S/N >100$. However, any inaccuracy here will have little effect on the estimated uncertainty 
in $RV$ for the majority of stars which are slow rotators since, at high values of $S/N$, 
the uncertainty of these stars is dominated by the constant term, $A$ in the expresssion for $S_{RV}$ (see eqns. 1 and 2).

The mean values of $B$ are 2.3\,km\,s$^{-1}$ for filter HR10 and 7.1\,km\,s$^{-1}$ for
filter HR21, compared with 5.0\,km\,s$^{-1}$ for HR15N i.e for spectra with similar $v\sin i$ and $S/N$ $RV$s 
estimated from spectra taken with HR10 are more precise. The temperature
dependence, illustrated in Fig.~5, is also different in detail. Figures 6a and b show the variation of $B$ 
with  $\log g$ and [Fe/H]. The trends are similar to the variation found for order sorting filter HR15N i.e
 $B$ is almost independent of gravity and changes only slowly with metallicity.

Parameter $A$~was determined in two ways. First, it was determined
using the measured data for the relatively small sample of long-term
repeats as described in Sect.~3.1. This gave values of
$A=0.18\pm0.02$\,km\,s$^{-1}$ for filter HR10 and $0.28\pm0.02$\,km\,s$^{-1}$ for
filter HR21. These compare with the predicted values of 0.22\,km\,s$^{-1}$ and
0.26\,km\,s$^{-1}$ inferred by scaling the $\delta_{pix}$ value determined for
filter HR15N by the ratio of their pixel sizes in km~s$^{-1}$.

Figures~10c and 10f shows the CDFs of the normalised uncertainty for short-
and long-term repeats using filters HR10 and HR21
respectively. In each case $S_{RV}$ is evaluated using the appropriate
theoretical  values of
$A$~and $C$~and the mean empirical value of $B$~determined for each
filter, and these are inserted into Eqns. 1 (for short-term repeats) or 2 (for
long-term repeats). Also shown is the CDF of a Student's t-distribution with
$\nu=6$ which, as for the HR15N data, appears to be an excellent
representation of the distribution due to short-term repeats. The data
are sparse for long-term repeats, but the distributions 
appear to have more extended tails, consistent with the idea that they
contain RV shifts due to binary systems. We do not have sufficient data
to test whether the uncertainty CDFs for long-term repeats within a
night are similar to those for short-term repeats, but we assume
that, like the HR15N data, this will the case for data taken with HR10
and HR21.
  
\section{Discussion and Summary}

We have shown that the normalisation functions given in Eqs. (1) --(4)
are reasonable descriptions of how uncertainties in $RV$ and $v \sin
i$ scale with $S/N$ and $v \sin i$. The recommended average parameters of $A$,
$B$, $C$ defining the scaling function for $RV$ are given in Table~3 for observations
performed with the three main instrumental configurations used for
GIRAFFE observations in the GES. Average values of $\alpha$, 
$\beta$, and $C$ that define the scaling function for $v\sin i$ are also given 
for filter HR15N. The uncertainties given by Eqns.~2 and 4
{\em are not} normally distributed; they have more extended tails. 
The uncertainty distribution for a given
observation of $RV$ is better represented by the value of Eq.~2 multiplied by
a normalised Student's t-distribution with $\nu=6$, whilst for $v \sin
i$ the uncertainty distribution can be approximated by Eq.~4 multiplied
by a normalised Student's t-distribution with $\nu=2$. 

Equations 2 and~4 decouple the influences of spectral type and the
spectrograph; $A$, $C$ and $\alpha$ are properties of the instrumental
setup, whilst $B$ and $\beta$ depend on the type of star observed. The dependence
on gravity, age and metallicity, over the range $-1<$[Fe/H]$<1$, is weak; but the temperature dependence becomes
strong for $T_{\rm eff} > 5200$\,K, such that $B$ and $\beta$ increase with $T_{\rm
  eff}$ and the precision worsens. This is presumably the result of a
decreasing number of strong, narrow lines in the spectra of hotter
stars. The temperature dependent $B$ and $\beta$ values are listed in
Table~3 and should be used in conjunction with the mean values of $A$,
$C$ and $\alpha$. There are insufficient observations of stars with 
$v \sin i>5$ km\,s$^{-1}$ using order-sorting
filters HR10 and HR21, so we cannot estimate $\beta$ for such observations. It
should also be noted that for reasons of sample size, the calibration
of $B$ and $\beta$ is limited to $3200 \leq T_{\rm eff} \leq  7200$\,K.

Parameter $A$ is between 0.22 and 0.26 km\,s$^{-1}$, dependent on
instrumental setup, and represents the best precision with which $RV$ can be
obtained from an individual GES spectrum with low rotational broadening and large $S/N$.
The origin of this term is unclear; it partly arises from uncertainties
in the wavelength calibration and the application of calibration
offsets from the ``simcal'' fibres or sky emission lines. However,
various tests have shown these cannot be entirely responsible and we
suspect there are additional contributions that may be associated with
movement of the fibres at the spectrograph slit assembly or target
mis-centering in the fibres combined with imperfect signal scrambling.

The analyses we present were derived from results in the iDR2 GES data
release and the coefficients in Table~3 are applicable to those data
and also to the more recent iDR3 update that used the same pipeline analysis.
The exact use of these results depends on the purpose of any particular
investigation. To estimate what approximates to a particular 
confidence interval on a $RV$ ($v \sin i$) value, the following
procedure is recommended:
\begin{enumerate}
\item Use the instrumental setup and an estimated stellar temperature
  (preferably from the GES analysis) to choose the appropriate values of $B$
  ($\beta$) and $C$ from Table~3; calculate $S_{RV,0}$ ($S_{v\sin
  i,0}$) from Eq.~1 (Eq.~3) using the measured $v \sin i$ and $S/N$.
\item Choose the $A$ ($\alpha$) value appropriate for the instrumental
  setup from Table 3 and calculate $S_{RV}$ ($S_{v\sin i}$) from Eq.~2 (Eq.~4). 
\item If combining results from repeated
  observations, these should be weighted using $S_{RV,0}^{-2}$ ($S_{v
    \sin i,0}^{-2})$ for short-term repeats (i.e. without the inclusion
  of the $A$ ($\alpha$) term), or $S_{RV}^{-2}$
  ($S_{v\sin i}^{-2}$) for long-term repeats.
\item For accurate
modelling of $RV$ data one should use $S_{RV}$ ($S_{v\sin i}$) multiplied by a
Student's t-distribution with $\nu=6$ ($\nu=2$) as a probability
distribution for the uncertainty. More crudely, a confidence interval
can be estimated by multiplying
$S_{RV}$ ($S_{v\sin i}$)  by the appropriate percentile point of a Student's
t-distribution with $\nu=6$ ($\nu=2$). For example, to estimate a 68.3
per cent error bar, multiply by 1.09 (1.32), or for a 95.4 per cent error bar
multiply by 2.51 (4.50). 
\end{enumerate}

Note that whilst the 68.3 per cent confidence intervals 
are quite close to the value
expected for a normal distribution with a standard deviation of
$S_{RV}$ ($S_{V\sin i}$), the 95.4 per cent confidence intervals 
are significantly larger
due to the broader tails of the Student's t-distributions. We do not
recommend extrapolating these estimates to even larger confidence
intervals since we have few data with
which to reliably constrain the distribution at these values. It seems
likely that at the conclusion of GES there will be sufficient data
(roughly 5 times as much) to significantly improve this situation. A
larger dataset will also allow us to study how $v \sin i$ precision
varies with $T_{\rm eff}$ and $\log g$ and between differing
observational setups.

\begin{acknowledgements}
RJJ wishes to thank the UK Science and Technology Facilities Council
for financial support.  Based on data products from observations made
with ESO Telescopes at the La Silla Paranal Observatory under programme
ID 188.B-3002. These data products have been processed by the Cambridge
Astronomy Survey Unit (CASU) at the Institute of Astronomy, University
of Cambridge, and by the FLAMES/UVES reduction team at
INAF/Osservatorio Astrofisico di Arcetri. These data have been obtained
from the Gaia-ESO Survey Data Archive, prepared and hosted by the Wide
Field Astronomy Unit, Institute for Astronomy, University of Edinburgh,
which is funded by the UK Science and Technology Facilities Council.
This work was partly supported by the European Union FP7 programme
through ERC grant number 320360 and by the Leverhulme Trust through
grant RPG-2012-541. We acknowledge the support from INAF and Ministero
dell' Istruzione, dell' Universit\`a' e della Ricerca (MIUR) in the
form of the grant "Premiale VLT 2012". The results presented here
benefit from discussions held during the Gaia-ESO workshops and
conferences supported by the ESF (European Science Foundation) through
the GREAT Research Network Programme.
\end{acknowledgements}

\bibliographystyle{aa} 
\bibliography{references} 

\appendix
\section{Variation of measurement precision with radial and projected
  rotation velocities}

We consider below how the measurement precision of $RV$ and $v\sin i$
scale with $S/N$ and $v\sin i$ for short-term repeats where there are
no changes in setup or wavelength calibration between observations. We
make the simplifying assumption that the precision in $RV$ scales as
$E_{RV} \propto W^{3/2}/(S/N)$ where $W$, is the FWHM of a Gaussian profile
representing the characteristic absorption line profile of the measured spectrum.
This approximate relation can be deduced from the results of Landman, Roussel-Dupre and Tanigawa (1982). 
These authors showed that, for the ideal case of a Gaussian line profile of amplitude $a$, mean value $m$, 
and standard deviation $s$, sampled using binned data with a uniform Gaussian noise of rms amplitude $\epsilon$ 
per bin, the statistical uncertainties in the estimated values of $m$ and $s$ are given by;

\begin{equation}
\sigma_m = s\left(\frac{4}{\pi}\right)^{1/4}\left(\frac{\Delta x}{s}\right)^{1/2}
\left(\frac{\epsilon}{a}\right)
\,\,	{\rm and}   \,\,  \sigma_s = \sigma_m 
\end{equation}
where $\Delta_x$ is the uniform bin width and $\Delta_x << s$.

In the present case $a$ varies with equivalent width, $(EW)$, of the characteristic absorption line as 
$a = (EW) h/\sqrt{2 \pi}\,s$, where $h$ is the amplitude of the contiuum. If the depth of the absortion line, 
$a$, is small compared to the continuum, $h$ then measurement uncertainty is $\epsilon \approx h/(S/N)$. 
Substituting these values in equation A1 using the relation $W = \sqrt{8\ln 2}\,s$ gives;

\begin{equation}
\sigma_m \propto \left(\frac{\Delta x^{1/2}}{EW}\right)\frac{W^{3/2}}{S/N}
\,	{\rm and}   \, 
\sigma_W \propto \left(\frac{\Delta x^{1/2}}{EW}\right)\frac{W^{3/2}}{S/N}
\end{equation} 

\subsection{Effect of $v\sin i$ on FWHM of the absorption line}

For a slowly rotating star, assuming that any sources of broadening other than rotation are much smaller than the intrinsic spectrograph resolution, the FWHM of an individual absorption line is
$W_0= \overline{\lambda}/R_{\lambda}$, where $\overline{\lambda}$ is the mean wavelength and $R_{\lambda}$
the resolving power of the spectrograph. For fast rotating stars the width of the spectral lines is increased by rotational broadening. Gray (1984) gives the rotational broadening
kernel as
\begin{equation}
	K(\lambda) =\frac{1}{\Lambda} \left(\frac{2(1-u)/\pi}{(1-u/3)}\sqrt{1-\left(\frac{\lambda}{\Lambda}\right)^2}
	          +\frac{u/2}{(1-u/3)}\left(1-\left(\frac{\lambda}{\Lambda}\right)^2\right)\right),
\end{equation}
where $\Lambda = \overline{\lambda}[v\sin i]/c$, $\lambda$ is wavelength (over the range $-\Lambda < \lambda < \Lambda$) and $u$ is the limb darkening coefficient.
  
Convolving a spectrum with this kernel increases the FWHM of individual lines  {\it approximately} as  $W \simeq \sqrt{W_0^2+(8\ln2)\lambda_{rms}^2}$  
where $\lambda_{rms}$~is the rms of the broadening kernel ($\lambda_{rms}^2=\int \lambda^2 K d\lambda$). Evaluating $\gamma_{rms}$~from Eq.~A3 gives;
\begin{equation} 
W = W_0\left(1+\left( \frac{v\sin i}{C}\right)^2\right)^{1/2}
\end{equation}
where $C = \left(\frac{1-u/3}{1-7u/15}\right)^{1/2}\frac{c}{R_{\lambda}\sqrt{2\ln2}}$

\subsection{Scaling of uncertainty in $RV$ and $v\sin i$}
To determine how the uncertainty in radial velocity, $E_{RV}$ scales
with $S/N$ and $v\sin i$ we assume $E_{RV} \propto \sigma_m$. For a
given spectra $\Delta x$  and $EW$ are  independent of $W$ and $S/N$ so that (from eqns. A2 and A4) $E_{RV}$ scales with $v\sin i$ and $S/N$ as
,  \begin{equation}
	S_{RV,0} = B\frac{\left (1+([v\sin i]/C)^2 \right)^{3/4}}{S/N}
\end{equation}

\noindent{where $B$ is an empirically determined constant and $C$ depends on $R_{\lambda}$ and 
$u$. A value of $u=0.6$ is used in this paper (Claret Diaz-Cordoves \& Gimenez 1995) giving $C=0.895 c/R_{\lambda}$.

The uncertainty in the estimated value of $v\sin i$ is determined from the uncertainty in the estimated absorption line width, $\sigma_W$ (eqns. A2 and A4) as; 

\begin{equation}
\sigma_{v\sin i} =\frac{C\,W\,\sigma_W}{W^2_0 \sqrt{W^2/W^2_0-1}}
\end{equation}
 Using this expession the uncertainty in the normalised value of $v\sin i$, ( $\propto \sigma_{v\sin i}/[v\sin i]$) scales with $v\sin i$ and $S/N$ as;

\begin{equation}
	S_{v\sin i,0} = \beta\frac{\left (1+([v\sin i]/C)^2 \right)^{5/4}}{(S/N)([v\sin i]/C)^2}\, ,
\end{equation}
where $\beta$ is an empirically determined constant.

\nocite{Butler1996a}
\nocite{Randich2013a}
\nocite{Cottaar2012a}
\nocite{Lardo2015a}
\nocite{Dufton2006a}
\nocite{Sacco2015a}
\nocite{Frasca2015a}
\nocite{Chandrasekhar1950a}
\nocite{Horne1986a}
\nocite{Gilmore2012a}
\nocite{Pasquini2002a}
\nocite{Koposov2011a}
\nocite{Jeffries2014a}
\nocite{Munari2005a}
\nocite{Gray1984a}
\nocite{Claret1995a}
\nocite{Landman1982a}
\nocite{Cottaar2014a}
\nocite{Lanzafame2015a}
\nocite{Luhman1999a}
\nocite{Luhman2007a}
\nocite{Jeffries2009b}
\nocite{Manzi2008a}
\nocite{Naylor2009a}
\nocite{Meynet1993a}
\nocite{Jeffries2005a}
\nocite{Strobel1991a}

%%%%%%%%%%%%%%%%%%%%%%%%%%%%%%%%%%%%%%%%%%%%%%%%%%%%%%%%%%%%%%%%

%%%%%%%%%%%%%%%%%%%%%%%%%%%%%%%%%%%%%%%%%%%%%%%%%%%%%%%%%%%%%%%%%

\label{lastpage}
\end{document}